\documentclass[journal=ancac3,manuscript=letter,layout=twocolumn]{achemso}

\usepackage[version=3]{mhchem} 
\usepackage{graphicx}
\usepackage{dcolumn}
\usepackage{bm}
\usepackage[mathlines]{lineno}
\usepackage{amssymb}
\usepackage{amsmath}
\usepackage{natbib}
\usepackage{soul,color}
\usepackage{xcolor,colortbl}

\newcommand*{\onlinecite}[1]{[\hspace{-1 ex} \nocite{#1}\citenum{#1}]}

\title{Chemical and Electronic Repair Mechanism of 
       Defects in MoS$_2$ Monolayers}

\author{Anja F\"{o}rster}
\affiliation{Physics and Astronomy Department,
             Michigan State University,
             East Lansing, Michigan 48824, USA}
\alsoaffiliation{Center for Advancing Electronics Dresden (cfaed),
             01062 Dresden, Germany}
\alsoaffiliation{Theoretical Chemistry,
             Technische Universit\"{a}t Dresden,
             01062 Dresden, Germany}

\author{Sibylle Gemming}
\affiliation{Center for Advancing Electronics Dresden (cfaed),
             01062 Dresden, Germany}
\alsoaffiliation{Helmholtz-Zentrum Dresden-Rossendorf,
             Institute of Ion Beam Physics and Materials Research,
             Bautzner Landstrasse 400,
             01328 Dresden, Germany }
\alsoaffiliation{Institute of Physics,
             Technische Universit\"{a}t Chemnitz,
             09107 Chemnitz, Germany}

\author{Gotthard Seifert}
\affiliation{Center for Advancing Electronics Dresden (cfaed),
             01062 Dresden, Germany}
\alsoaffiliation{Theoretical Chemistry,
             Technische Universit\"{a}t Dresden,
             01062 Dresden, Germany}
\alsoaffiliation{National University of Science and Technology,
             MISIS, Moscow, Russia}

\author{David Tom\'{a}nek}
\affiliation{Physics and Astronomy Department,
             Michigan State University,
             East Lansing, Michigan 48824, USA}
\email{tomanek@pa.msu.edu}

\keywords{transition metal dichalcogenides, 2D materials,
\textit{ab~initio} calculations, electronic structure, %
{defects}
\\}

\begin{document}


\begin{abstract}
Using {\em ab initio} density functional theory calculations, we
characterize changes in the electronic structure of MoS$_{2}$
monolayers introduced by missing or additional adsorbed sulfur
atoms. We furthermore identify the chemical and electronic
function of substances that have been reported to reduce the
adverse effect of sulfur vacancies in quenching photoluminescence
and reducing electronic conductance. We find that thiol-group
containing molecules adsorbed at vacancy sites may re-insert
missing sulfur atoms. In presence of additional adsorbed sulfur
atoms, thiols may form disulfides on the MoS$_{2}$ surface to
mitigate the adverse effect of defects.
\end{abstract}
\setlength\emergencystretch{1em}



There is growing interest in two-dimensional (2D) transition metal
dichalcogenide (TMD) semiconductors, both for fundamental reasons
and as potential components in flexible, low-power electronic
circuitry and for sensor applications
\cite{li2012fabrication,castellanos2013single,chang2013high}.
Molybdenum disulfide, MoS$_{2}$, is a prominent representative of
this class of TMDs. A free-standing, perfect 2D MoS$_{2}$
monolayer possesses a direct band gap of 1.88~eV at the $K$-point
in the Brillouin zone
\cite{ellis2011indirect,splendiani2010emerging}. Most commonly
used production methods for MoS$_{2}$ monolayers are chemical
vapor deposition (CVD) and mechanical
exfoliation of the layered bulk material~\cite{%
pachauri2013chemically,li2014preparation,varrla2015large}, as well
as sputter growth atomic layer deposition~\cite{Samassekou2017}
(ALD) of the precursor MoO$_3$ and subsequent conversion to the
disulfide under reducing conditions and at high
temperatures~\cite{Kastl2017,keller2017process}. A direct ALD
process using H$_2$S and MoCl$_5$\cite{Browning2015} or
Mo(CO)$_6$\cite{kwon2016comprehensive} is another possibility to
obtain MoS$_2$ monolayers. The CVD technique is probably best
suited for mass production, but the synthesized MoS$_{2}$ layers
lack in atomic perfection. The most common defects in these layers
are sulfur and molybdenum vacancies, as well as additional
adsorbed sulfur atoms
\cite{vancso2016intrinsic,addou2015impurities,liu2012growth,%
lee2012synthesis,zhan2012large,zhan_yuan_2015exploring}.
Eliminating or at least reducing the adverse effect of such
defects is imperative to improve the optoelectronic and transport
properties of TMDs.

In search of ways to mitigate the adverse effect of defects,
different methods have been suggested, including exposure of
MoS$_2$ to
superacids~\cite{{Javey15}}
or thiols\cite{yu2014towards,makarova2012selective}. %
{In the related MoSe$_2$ system, Se vacancies could be filled by S
atoms from an adjacent MoS$_2$ layer~\cite{Plochocka17}.} In the
present study, we focus on the reactions of thiols with defective
MoS$_2$ monolayers.

First, we characterize changes in the electronic structure of
MoS$_{2}$ monolayers introduced by missing or additional adsorbed
sulfur atoms using {\em ab initio} density functional theory (DFT)
calculations. We provide microscopic information about the
chemical and electronic function of thiols as a theoretical
background for the understanding of the successful use of thiols,
which have been reported to reduce the adverse effect of sulfur
vacancies in quenching photoluminescence and
to improve the electronic conductance of defective MoS$_2$. We
found that adsorbed thiols may re-insert missing sulfur atoms at
vacancy sites. %
{We also found that in presence of sulfur adatoms,} thiols will
form disulfides on the MoS$_{2}$ surface, which both mitigate the
adverse effect of defects.

\begin{figure}[t]
\includegraphics[width=1.0\columnwidth]{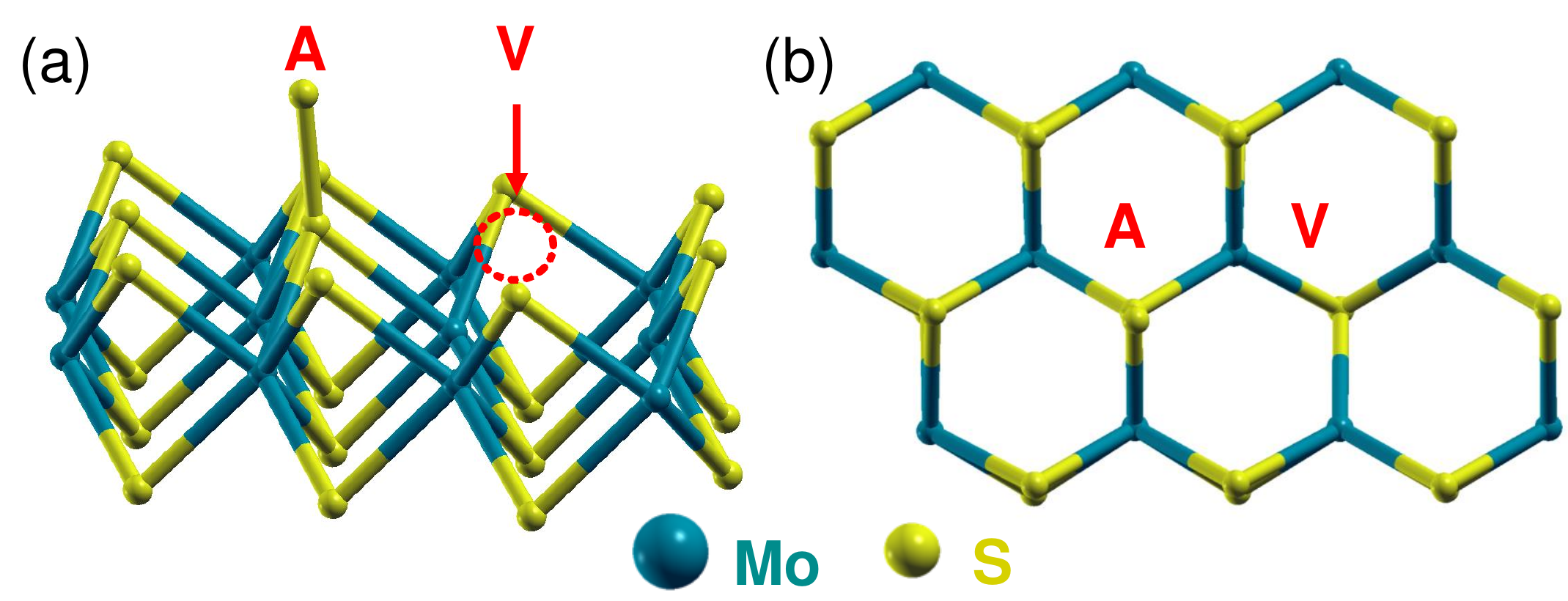}
\caption{(Color online) (a) Perspective and (b) top view of the
optimized geometry of an MoS$_{2}$ monolayer containing a sulfur
monovacancy (V) and a sulfur adatom (A).\label{fig1}}
\end{figure}

In Figure ~\ref{fig1} we display the structure of a defective
MoS$_{2}$ monolayer with a sulfur monovacancy (V) and an
additional sulfur adatom (A), since these defects are known to
significantly affect the electronic properties of MoS$_{2}$
\cite{noh2014stability}. The formation energy of the sulfur
vacancy is 2.71~eV and that of the sulfur adatom is 1.07~eV. %
{Consequently, the recombination energy of
a sulfur vacancy and a sulfur adatom
is $-1.89$~eV. In spite of the large energy gain, no spontaneous
healing will occur in a system with both defect types present due
to the high activation barrier of ${\approx}1.5$~eV for this
reaction. The listed defect formation energies are in agreement
with a study reporting the effect of various defect types on the
electronic structure of MoS$_2$~\cite{santosh2014impact}, and also
with a study of vacancy defects.\cite{seifert}}

\begin{figure}[t!]
\includegraphics[width=0.85\columnwidth]{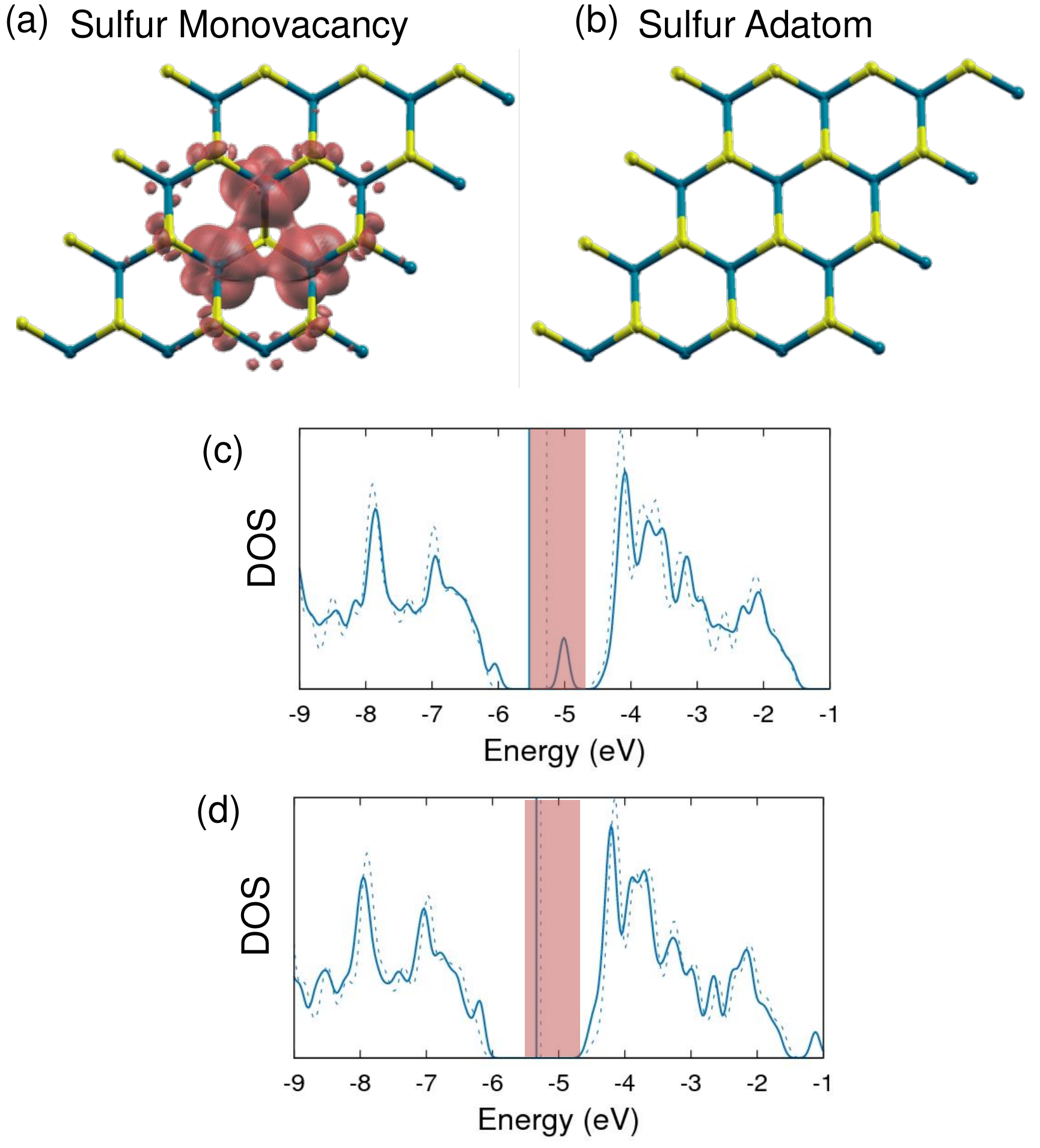}
\caption{(Color online) %
{Ball-and-stick models of (a) a sulfur vacancy defect and (b) a
sulfur adatom defect in an MoS$_2$ monolayer. Density of states
(DOS) of MoS$_2$ with (c) a vacancy and (d) an adatom defect. The
DOS and the position of the Fermi level are shown by solid blue
lines in defective lattices and by dotted blue lines in the
corresponding pristine lattices in (c) and (d). The DOS has been
convoluted by a Gaussian with a full-width at half maximum of
$0.1$~eV. The energy range of interest in the gap of the pristine
lattice is highlighted in red. The local density of states (LDOS),
representing the charge density associated with this highlighted
energy range, is represented by an isosurface and superposed to
the structure of a vacancy defect in (a) and an adatom defect in
(b). The isosurface value in the LDOS plots is
$0.003$~e/bohr$^3$.} \label{fig2}}
\end{figure}

Defects affect drastically the electronic structure in the
vicinity of the Fermi level. Setting apart the inadequacy of %
{DFT-PBE} calculations for quantitative predictions of band gaps,
we should note that in our computational approach with (large)
supercells and periodic boundary conditions, also defects form a
periodic array. In spite of their large separation, defect states
evolve into narrow bands that may affect the band structure of a
pristine MoS$_{2}$ monolayer. The effect of a sulfur monovacancy,
as well as that of a sulfur adatom, on the density of states (DOS)
of an MoS$_{2}$ monolayer around the band gap region is shown in
Figure~\ref{fig2}.

As seen in %
{Figure~\ref{fig2}c}, sulfur monovacancies introduce defect states
within the band gap and their superlattice shifts the DOS down by
0.16~eV %
{with respect to the pristine lattice.}
The defect states are localized around the vacancy %
{as seen in Figure~\ref{fig2}a}. The effect of a superlattice
of sulfur adatoms, addressed in Figure~\ref{fig2}b %
{and \ref{fig2}d, is to reduce the DFT band gap from $1.88$~eV to
$1.72$~eV,} in agreement with published
results~\cite{{noh2014stability},{santosh2014impact}}.

Defect sites play an important role as catalytically active
centers \cite{kwon2016comprehensive} and as sites for
functionalization reactions of 2D MoS$_{2}$
\cite{sim2015controlled}. Sulfur vacancies in particular are
considered to be important nucleation sites for a
functionalization with thiol molecules R-SH. The likely
possibility of an adsorbed thiol group transferring a sulfur atom
to the vacancy and thus repairing the defect is particularly
appealing. In this case, the detached hydrogen atom may reconnect
with the %
{remaining} R to form R-H and fill vacancy site of MoS$_{2}$ with
sulfur, as
\begin{equation}
 \textrm{R-SH} + \textrm{MoS}_{2}^{V} %
 \rightarrow %
 \textrm{R-H} + \textrm{MoS}_{2}, %
\label{eq:1}
\end{equation}
where MoS$_2^V$ denotes the MoS$_{2}$ layer with a sulfur vacancy.

An alternative reaction has been proposed to benefit from the STM
tip current in an STM study~\cite{makarova2012selective}. In the
first step of reaction (\ref{eq:2}), similar to reaction
(\ref{eq:1}), a hydrogen atom is removed from the thiol as its
sulfur atom fills the previous vacancy, determining the reaction
barrier for both reactions (\ref{eq:1}) and (\ref{eq:2a}). The
removed hydrogen atom will then form H$_2$ and desorb from the
MoS$_2$ surface. The rest R is still bound to the sulfur atom,
adsorbed at the sulfur vacancy site. The final assumption of the
proposed mechanism~\cite{makarova2012selective} is that the
R-groups are removed with the support of the STM tip, as
represented in reaction (\ref{eq:2b}).
\begin{subequations}
\label{eq:2}
\begin{align}
 \textrm{R-SH} + \textrm{MoS}_{2}^{V}%
 &\rightarrow %
 \frac{1}{2}\textrm{H}_{2} + \textrm{R-S-MoS}_{2}^{V}%
\label{eq:2a} \\
 \textrm{R-S-MoS}_{2}^{V} %
 &\xrightarrow{STM} %
 \textrm{R}{\bullet} + %
 \textrm{MoS}_{2} %
\label{eq:2b}
\end{align}
\end{subequations}
There is evidence in the literature supporting both reaction
(\ref{eq:1}) (\mbox{References~\onlinecite{yu2014towards}},
\mbox{\onlinecite{peterson1996ethanethiol}},
\mbox{\onlinecite{wiegenstein1999methanethiol}}) and reaction %
(\ref{eq:2})
(\mbox{Reference~\onlinecite{makarova2012selective}}).

The authors of
\mbox{Reference~\onlinecite{chen2016functionalization}} propose
yet another reaction (\ref{eq:3a}). Instead of the thiol molecules
repairing the sulfur vacancy, they form an adsorbed R-SS-R
disulfide at the surface of MoS$_2$ while releasing a hydrogen
molecule. We also considered the possibility that instead of
desorbing, the hydrogen molecule will fill the vacancy defect as
described in reaction (\ref{eq:3b}),
\begin{subequations}
\label{eq:3}
\begin{align}
 &\textrm{2 R-SH} + \textrm{MoS}_{2} \rightarrow %
 \textrm{R-SS-R} + \textrm{H}_{2} + \textrm{MoS}_{2} %
\label{eq:3a}\\
 &\textrm{2 R-SH} + \textrm{MoS}_{2}^{V} \rightarrow %
 \textrm{R-SS-R} + \textrm{H}_{2}\textrm{-MoS}_{2}^{V} %
\label{eq:3b}
\end{align}
\end{subequations}
Based on a previous study~\cite{benziger1985organosulfur} and the
observation of H$_{2}$S as well as H$_3$C=CH$_3$ during the
reaction of C$_{2}$H$_{5}$SH with bulk MoS$_2$ in
\mbox{Reference~\onlinecite{peterson1996ethanethiol}}, we also
considered a sulfur atom adsorbed on the MoS$_2$ surface,
identified as MoS$_2^A$, as the driving force for the observed
disulfide formation.

In this case, the reaction to form the disulfide R-SS-R is divided
into the following two steps. In reaction (\ref{eq:4a}), one thiol
reacts with the adatom to R-S-S-H and, in the follow-up reaction
(\ref{eq:4b}) with a second thiol, to R-S-S-R. An alternative
reaction with a sulfur vacancy following reaction (\ref{eq:4a}) is
also possible. Similar to reaction (\ref{eq:1}), the SH-group of
R-S-S-H can cure the vacancy defect, leading to the reduction of
R-S-S-H to the thiol R-S-H in Reaction (\ref{eq:4c}),
\begin{subequations}
\label{eq:4}
\begin{align}
\textrm{R-SH}+\textrm{MoS}_{2}^{A}\rightarrow
\textrm{R-SS-H}+\textrm{MoS}_{2}
\label{eq:4a}\\
\textrm{R-SH}+\textrm{R-S-S-H}\rightarrow\textrm{H}_{2}
\textrm{S}\textrm{\ensuremath{+}R-SS-R}
\label{eq:4b}\\
\textrm{R-SS-H}+\textrm{MoS}_{2}^{V}
\rightarrow\textrm{R-SH}+\textrm{MoS}_{2}%
\label{eq:4c}
\end{align}
\end{subequations}

\begin{figure}[!h]
\includegraphics[width=1.0\columnwidth]{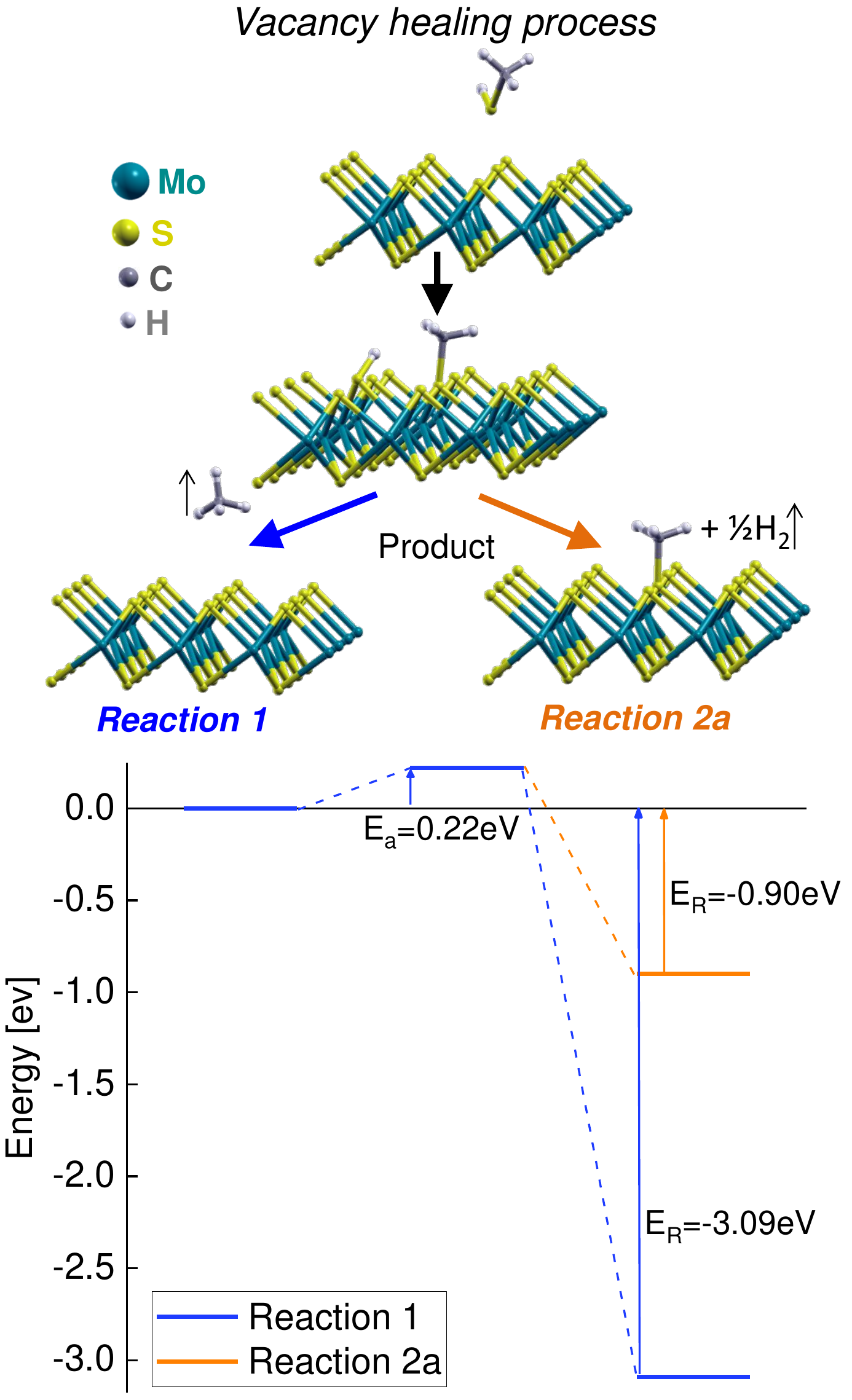}
\caption{(Color online) Reaction scheme for the sulfur vacancy
healing process caused by exposure of MoS$_{2}$ with vacancies to
CH$_{3}$SH. E$_R$ denotes the reaction energy and E$_a$ the
activation barrier. The same initial state can lead to two
different final states %
{{\em via}} the same transition state. The favorable reaction
(\ref{eq:1}), shown in dark blue, leads to a free CH$_{4}$
molecule. The energetics of reaction (\ref{eq:2a}) is displayed in
light orange. \label{fig3}}
\end{figure}

To better understand the above reaction mechanisms (\ref{eq:1}) --
(\ref{eq:4}), we performed DFT calculations to compare the energy
associated with the pathways of these reactions.
For the sake of easy understanding, we consider the
small methanethiol molecule CH$_{3}$SH as a representative of
thiols.

We limit our study of vacancy repair processes to reactions with
MoS$_{2}$ monolayers that contain one sulfur monovacancy per unit
cell. We analyze which reactions with thiols are favorable to
repair vacancy and adatom defects. Our results also unveil the
likely cause of apparent contradictions in the interpretation of
experimental results obtained by different researchers.

\section*{Results/Discussion}

\subsection*{Vacancy Repair}

The majority of published results indicate that thiol molecules
interacting with sulfur-deficient MoS$_{2}$ may fill in sulfur
atoms at the vacancy defect sites. Reaction pathways for the two
vacancy-healing reactions (\ref{eq:1}) and (\ref{eq:2a}), which
have been proposed in the
literature~\cite{{wiegenstein1999methanethiol},%
{peterson1996ethanethiol},{makarova2012selective}}, are sketched
in Figure~\ref{fig3}. We note that
reaction (\ref{eq:1}) has been studied in greater detail for a
different thiol~\cite{yu2014towards} and agrees with our findings
for the model compound H$_3$C-SH.

We find that reactions (\ref{eq:1}) and (\ref{eq:2a}) are both
exothermic and require crossing only a low activation barrier of
$0.22$~eV, since they share the same transition state shown in
Figure~\ref{fig3}. The larger energy gain $E_{R}=-3.09$~eV in
reaction (\ref{eq:1}) in comparison to $-0.90$~eV in reaction
(\ref{eq:2a}) suggests that the former reaction is
thermodynamically preferred.

\begin{figure}[h]
\includegraphics[width=0.8\columnwidth]{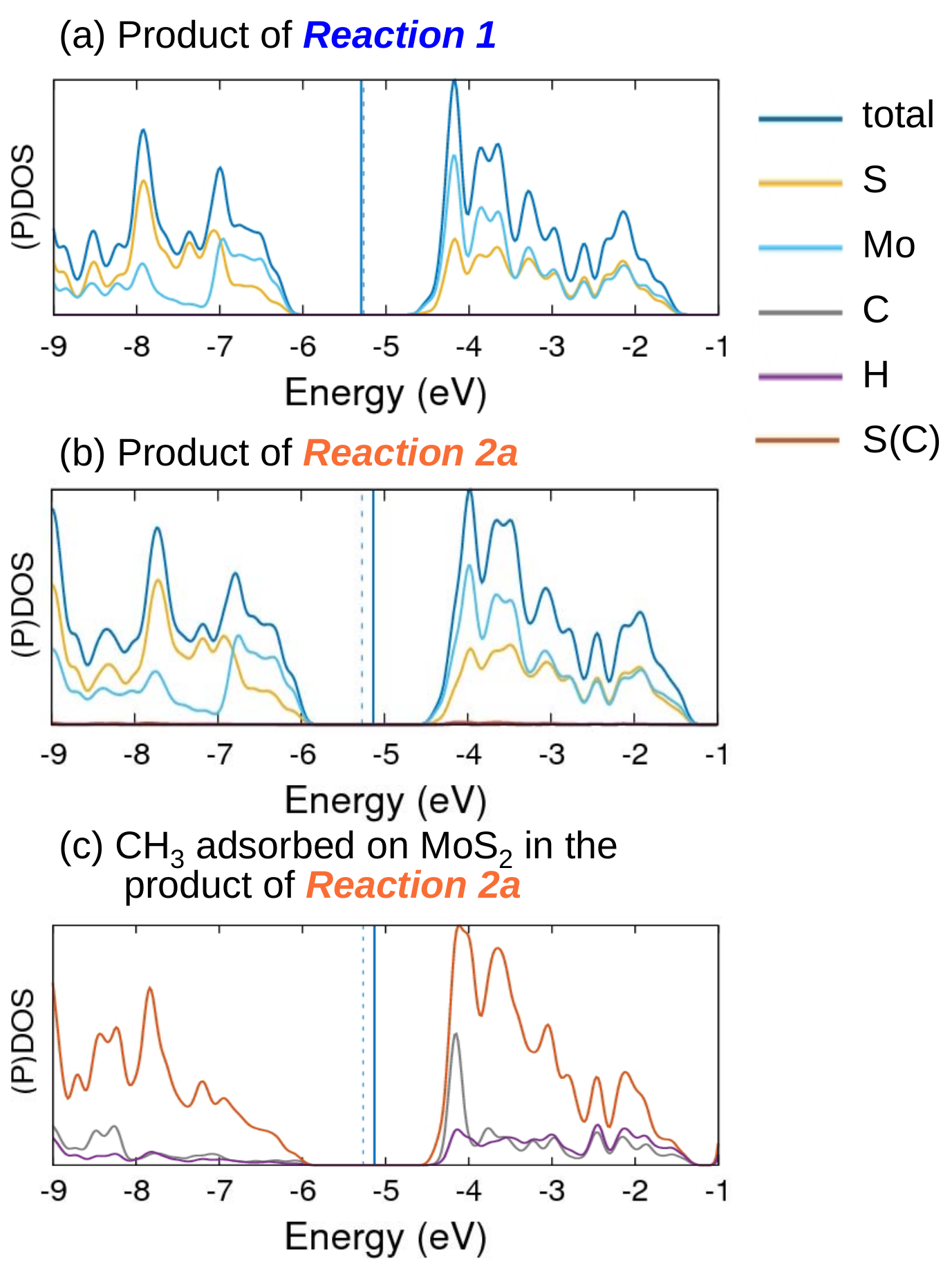}
\caption{(Color online) %
{Electronic structure of products of the vacancy healing process
shown in Figure~\protect\ref{fig3}. The total density of states
(DOS) and partial densities of states (PDOS) of the products of
reactions (a) (\ref{eq:1}) and (b) (\ref{eq:2a}). (c) PDOS of the
CH$_3$-molecule bonded to a sulfur atom in an MoS$_2$ monolayer.
The PDOS of the sulfur atom connected to C in CH$_3$, S(C), is
shown by the brown line. All DOS and PDOS functions have been
convoluted by a Gaussian with a full-width at half maximum of
$0.1$~eV. The position of the Fermi level is shown by solid blue
lines in defective lattices and by dotted blue lines in the
corresponding pristine lattices.} \label{fig4}}
\end{figure}
\begin{figure}[t]
\includegraphics[width=0.8\columnwidth]{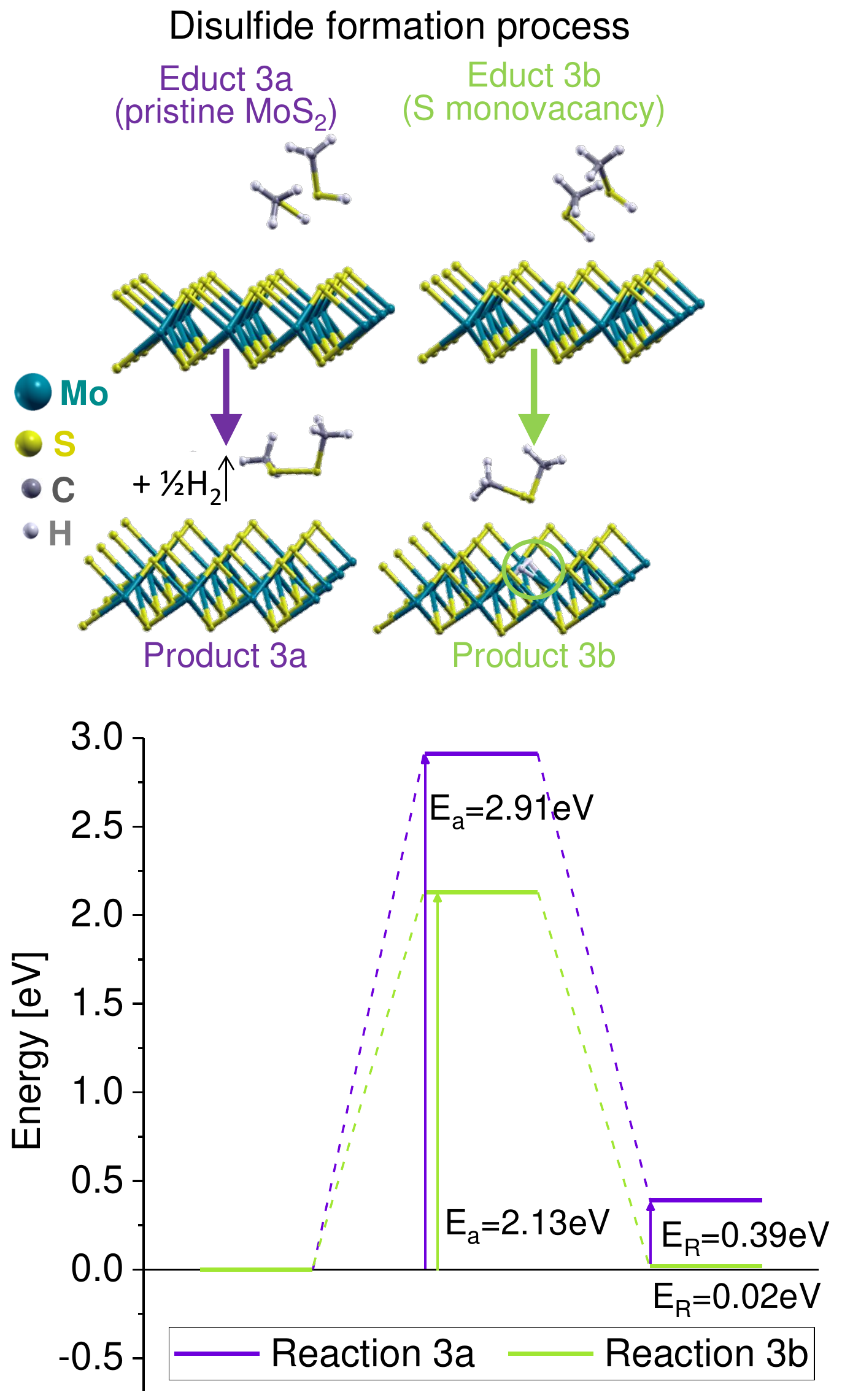}
\caption{(Color online) Reaction scheme of the disulfide formation
process involving exposure of an MoS$_{2}$ monolayer to two
CH$_{3}$SH molecules. %
{E$_R$ denotes the reaction energy and E$_a$ the activation
barrier.} Reaction (\ref{eq:3a}), shown in dark purple, occurs on
pristine MoS$_{2}$. Reaction (\ref{eq:3b}), shown in light green,
occurs on a sulfur-deficient MoS$_{2}$ substrate. \label{fig5}}
\end{figure}

Figure~\ref{fig4}a shows the DOS %
{and partial densities of states (PDOS), projected on individual
atoms,} of the product of
reaction (\ref{eq:1}). %
{Figure~\ref{fig4}}b provides the corresponding information for
reaction (\ref{eq:2a}), %
{and Figure~\ref{fig4}c provides a detailed view of the PDOS for
the CH$_3$-group and the connected sulfur atom.} In both cases,
the defect states associated with sulfur monovacancies have been
removed. In the final state of reaction (\ref{eq:1}) the DOS is
completely restored to the undamaged state of the semiconductor.
For reaction (\ref{eq:2a}), on the other hand, the Fermi level is
shifted to the lower edge of the conduction band %
{due to the CH$_3$-group}. Therefore, only the preferred repair
reaction (\ref{eq:1}) leads to both an electronic and a chemical
repair of MoS$_{2}$.

\subsection*{Disulfide Formation}

A different reaction scenario (\ref{eq:3a}) has been proposed in
\mbox{Reference~\onlinecite{chen2016functionalization}},
suggesting that disulfides are formed when thiols interact with
MoS$_{2}$. We investigated the MoS$_2$ surface both in its
pristine state and in presence of sulfur vacancies to clarify the
differences in the catalytic potential for disulfide formation.
Figure~\ref{fig5} illustrates reaction (\ref{eq:3a}) on pristine
MoS$_2$ and reaction (\ref{eq:3b}) on a sulfur-deficient MoS$_2$
substrate. The schematic reaction profile indicates that both
reactions involve significant activation barriers. As seen in
Figure~\ref{fig5}, reaction (\ref{eq:3a}) is endothermic with a
reaction energy of E$_R$=0.39~eV and involves a high activation
barrier of 2.91~eV. Reaction (\ref{eq:3b}) near a sulfur
monovacancy is only slightly endothermic, with a reaction energy
E$_R$=0.02~eV, and involves a somewhat lower activation barrier of
2.13~eV. Thus, reaction (\ref{eq:3b}) is energetically more
favorable than reaction (\ref{eq:3a}).

The near-neutral reaction energy of reaction (\ref{eq:3b}) can be
explained by the Kubas interaction of transition metal
$\eta^{2}$-H$_2$-complexes.\cite{kubas1980five,gordon2010perspectives}
It means that the excess H$_2$ molecule in the product of reaction
(\ref{eq:3b}), which is attached to an Mo atom at the sulfur
vacancy site and indicated by a circle in Figure~\ref{fig5}, still
retains the H-H bond character. According to
\mbox{Reference~\onlinecite{Skipper2012}}, the Kubas interaction
energy is in the range of $0.2-0.4$~eV, in agreement with the
reaction energy difference of 0.37~eV between reactions
(\ref{eq:3a}) and (\ref{eq:3b}).

\begin{figure}[t]
\includegraphics[width=0.95\columnwidth]{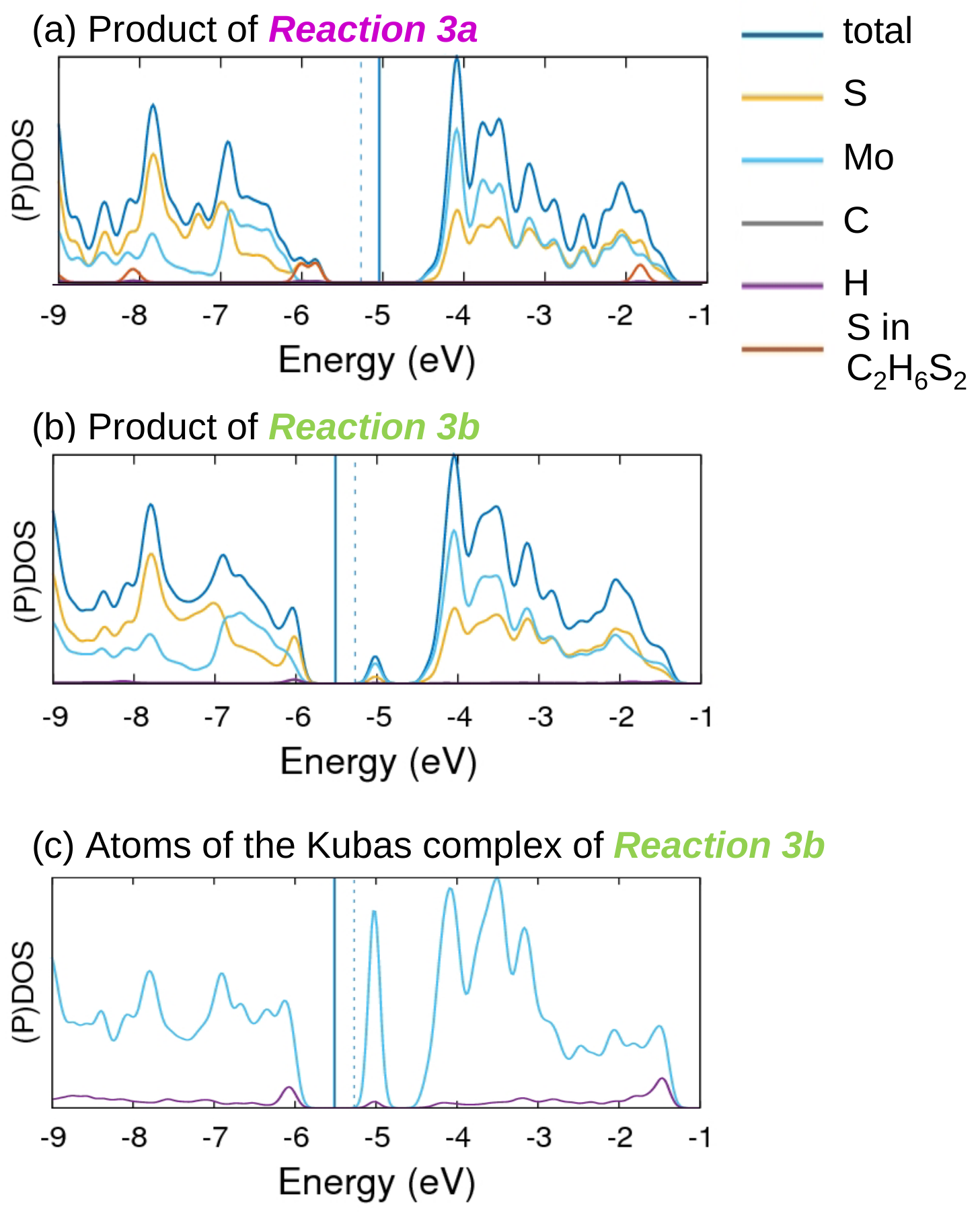}
\caption{(Color online) %
{Electronic structure of products of the disulfide formation
process shown in Figure~\protect\ref{fig5}. The total density of
states (DOS) and partial densities of states (PDOS) of the
products of reactions (a) (\ref{eq:3}a) and (b) (\ref{eq:3}b). (c)
PDOS of the 3 Mo atoms and the H$_2$ molecule attached to the
vacancy that form the Kubas complex in the product of reaction
(\ref{eq:3}b). All DOS and PDOS functions have been convoluted by
a Gaussian with a full-width at half maximum of $0.1$~eV. The
position of the Fermi level is shown by solid blue lines in
defective lattices and by dotted blue lines in the corresponding
pristine lattices.} %
\label{fig6}}
\end{figure}
\begin{figure}[t]
\includegraphics[width=0.8\columnwidth]{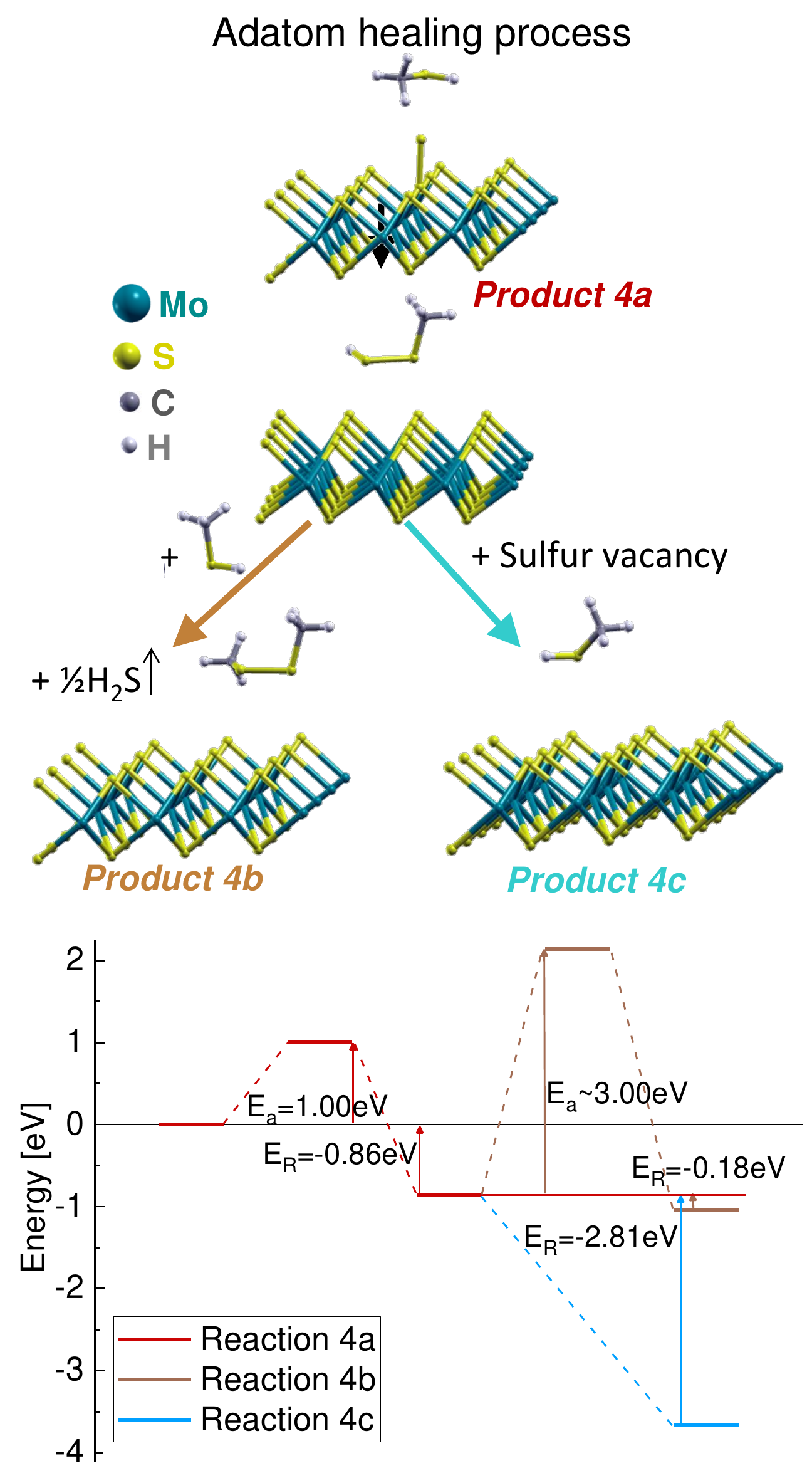}
\caption{(Color online) %
{Reaction scheme of the adatom healing process that starts with
reaction (\ref{eq:4a}) and leads to} disulfide formation in
presence of extra sulfur atoms on MoS$_2$, shown in dark red.
Subsequent ligand exchange reaction (\ref{eq:4b}) is shown in
brown. Alternative subsequent vacancy
repair reaction (\ref{eq:4c}) is shown in light blue. %
{E$_R$ denotes the reaction energy and E$_a$ the activation
barrier.} %
\label{fig7}}
\end{figure}

The Kubas interaction also reduces the reaction barrier and degree
of endothermicity considerably. Nevertheless, reactions
(\ref{eq:3a}) and (\ref{eq:3b}) are not competitive in comparison
with the strongly exothermic reaction (\ref{eq:1}) with
E$_R=-3.09$~eV in thermodynamic equilibrium.

The %
{PDOS functions} characterizing the products of reactions
(\ref{eq:3a}) and
(\ref{eq:3b}), %
{visualized in Figure~\ref{fig5},} are shown in Figure~\ref{fig6}.
The product of reaction (\ref{eq:3b}) still contains a defect
state in the gap region, indicating that the chemisorbed H$_2$
molecule is incapable of
electronically repairing the effect of the sulfur vacancy. %
{This is seen in the PDOS of the Mo atoms of the Kubas complex
surrounding the vacancy defect in Figure~\ref{fig6}c.} The product
of reaction (\ref{eq:3a}), on the other hand, shows no indication
of a defect state, since the vacancy-free MoS$_2$ monolayer is not
affected much by the %
{physisorbed}
disulfide, %
{as seen in the PDOS of Figure~\ref{fig6}a}.

We can thus conclude that the disulfide formation reaction
(\ref{eq:3a}), suggested in
\mbox{Reference~\onlinecite{chen2016functionalization}}, is
endothermic. The alternative reaction (\ref{eq:3b}) on a
sulfur-deficient MoS$_2$ substrate displays a lower activation
barrier and an end-product stabilized by the Kubas interaction,
but is still weakly endothermic and thus unlikely. In the
following, we propose an alternative pathway towards disulfide
formation.

\begin{figure}[h]
\includegraphics[width=1.0\columnwidth]{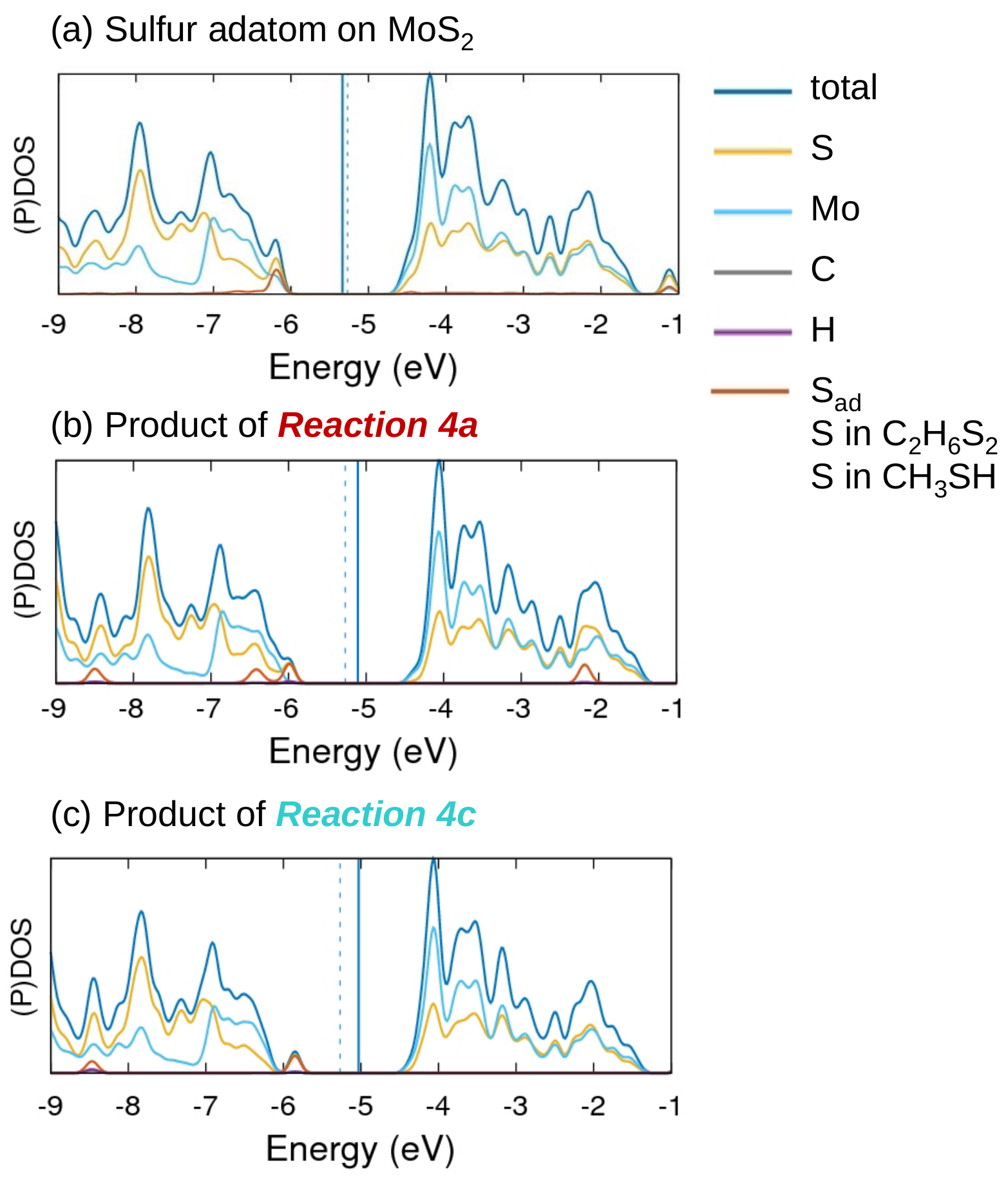}
\caption{(Color online) %
{Electronic structure of products of the adatom healing process
shown in Figure~\protect\ref{fig7}. The total density of states
(DOS) and partial densities of states (PDOS) of (a) MoS$_2$ with a
sulfur adatom (without CH$_3$SH), (b) the product of reaction
(\ref{eq:4a}), and (c) the product of reaction (\ref{eq:4c}). All
DOS and PDOS functions have been convoluted by a Gaussian with a
full-width at half maximum of $0.1$~eV. The position of the Fermi
level is shown by solid blue lines in defective lattices and by
dotted blue lines in the corresponding pristine lattices.} %
\label{fig8}} %
\end{figure}

\subsection*{Adatom Repair}

The postulated alternative reaction requires extra sulfur atoms
adsorbed on the MoS$_2$ surface, which act as nucleation sites for
the disulfide formation. The reaction leading to the formation of
disulfide R-SS-R in presence of sulfur adatoms consists of two
steps, described by reactions (\ref{eq:4a}) and (\ref{eq:4b}), as
well as the alternative reaction (\ref{eq:4c}) following reaction
(\ref{eq:4a}), as shown in Figure~\ref{fig7}.

In reaction (\ref{eq:4a}), a CH$_3$SH molecule interacts with the
reactive sulfur adatom to methylhydro-disulfide (CH$_3$SSH),
releasing $-0.86$~eV due to the formation of a stable disulfide
bond. The estimated activation barrier for this reaction is close
to 1~eV, which is considerably lower than the values for the
corresponding reactions (\ref{eq:3a}) and (\ref{eq:3b}) in absence
of an extra sulfur adatom. %

{Electronic structure changes during the adatom healing process
are displayed in Figure~\ref{fig8}.} The DOS of the product of
reaction (\ref{eq:4a}), shown in Figure~\ref{fig8}b, shows no
defect-related states in the band gap, indicating chemical and
electronic repair of the sulfur adatom defect %
{that is seen in Figure \ref{fig8}a}.

In the subsequent reaction (\ref{eq:4b}), shown in
Figure~\ref{fig7}, a second CH$_3$SH molecule interacts with the
methylhydro-disulfide CH$_3$SSH, leading to the exchange of the
hydrogen atom with a methyl group and formation of hydrogen
sulfide (H$_2$S) as a side product. This reaction is mildly
exothermic, with an overall reaction energy of $-0.18$~eV. Even
though the combined reaction (\ref{eq:4a}) and (\ref{eq:4b}) for
the formation of CH$_3$SSCH$_3$ is strongly exothermic with a net
energy gain of $-1.05$~eV, the activation barrier for the ligand
exchange in reaction (\ref{eq:4b}) is prohibitively high with
$E_a{\approx}+3$~eV, which essentially suppresses the formation of
CH$_3$SSCH$_3$ following reaction (\ref{eq:4a}).

Therefore, we investigated reaction (\ref{eq:4c}) as an
alternative follow-up process to reaction (\ref{eq:4a}). In
reaction (\ref{eq:4c}), the CH$_3$SSH molecule interacts with a
nearby sulfur vacancy defect. This reaction is similar to the
vacancy healing reaction (\ref{eq:1}) and consequently is strongly
exothermic with a reaction energy of $-2.81$~eV. Reaction
(\ref{eq:4c}) is barrier-free and thus occurs spontaneously. As
seen in in Figure~\ref{fig8}c, describing the product of reaction
(\ref{eq:4c}), the defect-related state above $E_F$ has been
removed from the DOS. This means that following the adatom repair
and disulfide formation reaction (\ref{eq:4a}), reaction
(\ref{eq:4c}) will take place in case that also sulfur vacancies
are present. The two reactions will thus heal both vacancy and
adatom defects.

Our above considerations offer an attractive explanation why
disulfide formation was observed in
\mbox{Reference~\onlinecite{chen2016functionalization}}, %
{but not in %
\mbox{References~\onlinecite{yu2014towards}},
\mbox{\onlinecite{makarova2012selective}},
\mbox{\onlinecite{peterson1996ethanethiol}} and
\mbox{\onlinecite{wiegenstein1999methanethiol}}. %
Initially, reactions (\ref{eq:1}) and (\ref{eq:4a}) plus
(\ref{eq:4c}) have taken place in all samples that contained
vacancies. Vacancy healing as primary outcome of reactions
reported in \mbox{References~\onlinecite{yu2014towards}},
\mbox{\onlinecite{makarova2012selective}},
\mbox{\onlinecite{peterson1996ethanethiol}} and
\mbox{\onlinecite{wiegenstein1999methanethiol}} could likely be
achieved due to an abundance of vacancies in the samples used. We
may speculate that the MoS$_{2}$ sample of
\mbox{Reference~\onlinecite{chen2016functionalization}} contained
more sulfur adatoms than sulfur vacancies. In that case, all
vacancy defects could be repaired, but some adatom defects were
left unrepaired in the sample of
\mbox{Reference~\onlinecite{chen2016functionalization}}. At this
point, lack of vacancy defects would block reactions (\ref{eq:1})
and (\ref{eq:4c}). The only viable reaction was (\ref{eq:4a}),
which repaired adatom defects, leaving a pristine MoS$_2$ surface
behind with disulfide as a by-product .} This speculative
assumption is also consistent with the observation that the
electronic structure of MoS$_2$ has remained unaffected by the
reaction leading to the formation of
disulfide.\cite{chen2016functionalization}


\section*{Conclusions}

We studied three different reaction paths of thiols, represented
by methanethiol (CH$_{3}$SH), %
{with a defective 2D MoS$_{2}$ monolayer. We showed that the
repair of sulfur monovacancies by adsorbed CH$_{3}$SH is an
exothermic reaction releasing up to $3$~eV. In another possible
reaction between CH$_{3}$SH and MoS$_{2}$, leading to the
formation of disulfide, we found that presence of sulfur vacancies
lowers the reaction barrier due to the Kubas interaction at the
defect site. The corresponding reaction involving MoS$_2$ with
sulfur adatoms instead of vacancies, on the other hand, leads to
disulfide formation and releases about $0.9$~eV. In the presence
of sulfur vacancies, the formed disulfides will immediately reduce
to thiols while simultaneously healing the vacancy defect. We can
therefore conclude that, regardless of interim disulfide
formation, thiols always lead to a chemical repair of available
sulfur vacancies by filling-in the missing sulfur atoms and
consequently eliminating vacancy-related defect states in the
gap.}

\section*{Methods/Theoretical}

To obtain insight into the reaction processes, we performed DFT
calculations using the SIESTA code~\cite{soler2002siesta}. %
{We used {\em ab initio} Troullier-Martins
pseudopotentials~\cite{Troullier91} and the Perdew-Burke-Ernzerhof
(PBE) exchange-correlation functional~\cite{perdew1996generalized}
throughout the study. Except for sulfur, all pseudopotentials used
were obtained from the on-line resource in
\mbox{Reference~\onlinecite{Pseudopotential}}.
The pseudopotential of sulfur has been generated without core
corrections using the ATM code in the SIESTA suite and the
parameters listed in
\mbox{Reference~\onlinecite{Pseudopotential}}. All
pseudopotentials were tested against atomic all-electron
calculations. We used a double-$\zeta$ basis set including
polarization orbitals (DZP) to represent atoms in crystal
lattices, $140$~Ry as the mesh cutoff energy for the Fourier
transform of the charge density, and $0$~K for the electronic
temperature. We used periodic boundary conditions with large
supercells spanned by the lattice vectors $\vec{a}_1=(12.84, 0.00,
0.00 )$~{\AA}, $\vec{a}_2=(6.42, 11.12, 0.00)$~{\AA},
$\vec{a}_3=(0.00, 0.00, 22.23)$~{\AA} to represent pristine and
defective 2D MoS$_{2}$ lattices. The unit cells of defect-free
MoS$_{2}$ contained 16 molybdenum and 32 sulfur atoms, and were
separated by a vacuum region of ${\approx}15$~{\AA} normal to the
layers. The Brillouin zone was sampled by a $4{\times}4{\times}1$
$k$-point grid~\cite{monkhorst1976special} and its equivalent in
larger supercells. }

{The above input parameters were found to guarantee convergence.
In particular, we found that using the larger triple-$\zeta$
polarized (TZP) instead of the DZP basis and increasing the mesh
cutoff energy affected our total energy differences by typically
less than $0.01$~eV. We furthermore validated the {\em ab initio}
pseudopotential approach used in the SIESTA code by comparing to
results of the all-electron SCM-Band code~\cite{SCMBand} and found
that energy differences obtained using the two approaches differed
typically by less than $0.3$~eV. }

All geometries have been optimized using the conjugate gradient
method \cite{CGmethod}, until none of the residual
Hellmann-Feynman forces exceeded $10^{-2}$~eV/{\AA}. In addition
to the default density matrix convergence, we also demanded that
the total energy should reach the tolerance of
${\lesssim}10^{-4}$~eV. %
{To eliminate possible artifacts associated with local minima, we
verified initial and final state geometries by performing
canonical molecular dynamics {(MD)} simulations using the
NVT-Nos\'{e} thermostat with $T=273.15$~K and $1$~fs time steps.}

{Due to the complexity of the reaction energy hypersurface and the
large number of relevant degrees of freedom, approaches such as
the nudged elastic band, which are commonly used to determine the
reaction path including transition states, turned out to be
extremely demanding on computer resources. We focussed on
transition states only and initiated our search by running
canonical MD simulations starting from a set of educated guesses
for the geometry. Following the atomic trajectories, we could
identify a saddle point in the energy hypersurface, where all
forces acting on atoms vanished, and postulated this point in
configurational space as a transition state. To confirm this
postulate, we ran MD simulations starting at a slightly altered
geometry of the postulated transition state. We concluded that the
postulated transition state is indeed the real transition state
once all trajectories reached either the initial (educt) or the
final (product) state. The activation barrier was determined by
the energy difference between the initial and the transition
state. }
\\


{\noindent\bf Author Information}\\

{\noindent\bf Corresponding Author}\\
$^*$E-mail: {\tt tomanek@pa.msu.edu}

{\noindent\bf Notes}\\
The authors declare no competing financial interest.

\begin{acknowledgement}
We thank Jie Guan and Dan Liu for useful discussions and Garrett
B. King for carefully checking the bib\-lio\-graphy. This study
was supported by the NSF/AFOSR EFRI 2-DARE grant number
\#EFMA-1433459. Computational resources have been provided by the
Michigan State University High Performance Computing Center %
{and the Center of Information Services and High Performance
Computing (ZIH) at TU Dresden.} AF, SG and GS acknowledge funding
from the Center for Advancing Electronics Dresden (cfaed). AF
especially acknowledges the cfaed Inspire Grant. SG acknowledges
funding from the Initiative and Networking Funds of the President
of the Helmholtz Association  %
{{\em via}} the W3 programme.
\end{acknowledgement}


\providecommand{\latin}[1]{#1} \makeatletter \providecommand{\doi}
  {\begingroup\let\do\@makeother\dospecials
  \catcode`\{=1 \catcode`\}=2\doi@aux}
\providecommand{\doi@aux}[1]{\endgroup\texttt{#1}} \makeatother
\providecommand*\mcitethebibliography{\thebibliography} \csname
@ifundefined\endcsname{endmcitethebibliography}
  {\let\endmcitethebibliography\endthebibliography}{}

\end{document}